\documentstyle[epsf,12pt]{article}

\begin{document}

\renewcommand{\theequation}{\thesection.\arabic{equation}}
\newcommand{\reseteqnum}{\setcounter{equation}{0}}

\title{
\hfill
\parbox{4cm}{\normalsize KUNS 1477\\HE(TH)~97/17\\hep-th/9711090}\\
\vspace{2cm}
D-branes on Orbifolds and Topology Change
\vspace{1.5cm}}
\author{Tomomi Muto\thanks{e-mail address:
\tt muto@gauge.scphys.kyoto-u.ac.jp}\\
{\normalsize\em Department of Physics, Kyoto University}\\
{\normalsize\em Kyoto 606-01, Japan}}
\date{\normalsize }
\maketitle
\vspace{1cm}

\begin{abstract}
\normalsize

We consider D-branes on an orbifold ${\bf C}^3/{\bf Z}_n$ and investigate the
moduli space of the D-brane world-volume gauge theory by using toric geometry
and gauged linear sigma models. For $n=11$, we find that there are five phases,
which are topologically distinct and connected by flops to each other. We
also verify that non-geometric phases are projected out for $n=7,9,11$ cases
as expected. Resolutions of non-isolated singularities are also investigated.

\end{abstract}

\newpage
\section{Introduction}

In string theory, it had been thought that
standard concepts of space-time would break
down at the scale $\sqrt {\alpha '}$
since it is the scale of probes, i.e.
fundamental strings.
In \cite{DKPS}, however,
it was argued that the structure of space-time
on sub-stringy scales can be probed by D-branes.
Space-time appearing in the D-brane approach
has very different features from that
probed by fundamental strings.
First of all, space-time coordinates
are promoted to non-commuting matrices,
and usual space-time emerges from
moduli space of D-brane world-volume
gauge theory.
So it is interesting to investigate
space-time by using D-branes as probes
and compare it with space-time
probed by fundamental strings.
Investigations toward this direction
were made in \cite{DM,JM,DGM}, in which
D-branes on orbifolds were studied.
In particular,
three dimensional orbifolds in \cite{DGM}
serve as local descriptions of singularities
in Calabi-Yau manifolds
\footnote{In \cite{Mohri}, orbifold singularities
of Calabi-Yau fourfolds were investigated
along the lines of \cite{DGM}.}.

Investigations on the moduli space of Calabi-Yau manifolds
were made in \cite{AGM,Witten} based on fundamental strings.
It was shown that the moduli space has a rich phase structure.
It includes topologically distinct Calabi-Yau
spaces and non-geometric phases
such as orbifold and Landau-Ginzburg phases.
Topologically distinct Calabi-Yau phases
are connected by flops.
A flop is achieved by a sequence of operations:
first blowing down some homologically
nontrivial cycle ${\bf CP}^1$
and then blowing up another ${\bf CP}^1$.
In this process, Hodge numbers do not change
but more subtle topological indices
such as intersection numbers
among homology cycles change.
In the course of the flop,
the space becomes singular due to the shrinking
of some ${\bf CP}^1$.
As a conformal field theory, however, this process
occurs smoothly.
Possibility of the smooth topology change is
demonstrated by using mirror symmetry in \cite{AGM}.
In the gauged linear sigma model approach\cite{Witten},
the singularity is avoided by giving non-zero
theta angle.
As for the non-geometric phases,
it is argued that they can also be interpreted
geometrically by analytic continuation to
Calabi-Yau phases, although part of the Calabi-Yau
manifold
has been shrunk to string or sub-stringy scales
\cite{AGM2}.

In \cite{DGM}, the moduli space of Calabi-Yau
manifolds was investigated by using
D-branes as probes.
They considered D-branes in typeII string theory
on orbifolds ${\bf C}^3/{\bf Z}_3$ and ${\bf C}^3/{\bf Z}_5$.
It gives $U(1)^n$ gauge theory with
Fayet-Iliopoulos D-terms coming from
twisted sectors of closed strings.
As the coefficients of the Fayet-Iliopoulos
D-terms change, the moduli space of the gauge
theory changes.
A priori, it may seem that a rich phase structure
arises as in \cite{AGM,Witten}.
However it is shown that only Calabi-Yau
phases are allowed and non-geometric
phases are projected out.
This result matches the analytically continued
picture of the moduli space in \cite{AGM2}.
It is also consistent with the study of the moduli
space of Calabi-Yau compactifications
in M theory\cite{Witten2},
which can be thought of as
strong coupling limit of type IIA theory.

To proceed the comparison between the moduli space
of Calabi-Yau spaces probed by fundamental strings
and that probed by D-branes,
it is important to investigate topology changing
process in the D-brane approach.
In this paper we present an explicit example
in which the moduli space includes topologically
distinct Calabi-Yau phases connected by flops
based on D-brane world-volume gauge theory.

The organization of this paper is as follows.
In section 2, we review flops in terms of toric
geometry and gauged linear sigma models which are
necessary to the analyses in the following sections.
In section 3, we explain D-branes on orbifold
${\bf C}^3/{\bf Z}_n$.
In section 4, we first review the work \cite{DGM},
which treats $n=3,5$ cases.
We then consider $n=7,9,11$ cases
and explicitly check that non-geometric phases are
projected out.
In the $n=11$ case, we present
the model in which there are five
topologically distinct phases connected by flops.
In section 5, we consider D-brane on orbifolds with
non-isolated singularities.
Section 6 contains discussion.

\section{Topology change in toric geometry}
\reseteqnum

In this section, we review toric varieties
and physical realization of toric varieties
in terms of gauged linear sigma models
emphasizing topology changing process.
For details, see \cite{AGM,Witten,AG,MP}.

A complex $d$-dimensional toric variety
is a space which contains algebraic
torus $({\bf C}^*)^d$ as a dense open subset.
A toric variety is determined by a combinatorial
data $\Delta$ called a fan,
so we denote it by $V_\Delta$.
A fan $\Delta$ is a collection
of strongly convex rational polyhedral cones
in ${\bf R}^d$ with apex at the origin.
To be a fan it must have the property that
(1) any two members of the collection
intersect in a common face,
(2) for each member of $\Delta$ all its face
are also in $\Delta$.

$V_\Delta$ can be expressed in the form
$({\bf C}^k-F_\Delta)/({\bf C}^*)^{k-d}$,
where $k$, $F_\Delta$ and the action of
$({\bf C}^*)^{k-d}$ on ${\bf C}^k$
are determined by $\Delta$ as follows.

\noindent
1. Let $\vec{n}_1,\vec{n}_2,\ldots,\vec{n}_k$
be the integral generators of the one dimensional
cones in $\Delta$.
Then associate a homogeneous coordinate $p_i$ of
${\bf C}^k$ with each vector
$\vec{n}_i$.

\noindent
2. Define $F_\Delta$, a subset of ${\bf C}^k$, by
\begin{equation}
F_\Delta=\bigcap_{\sigma \in \Delta}
\{(p_1,p_2,\ldots,p_k) \in {\bf C}^k;
\prod_{\vec{n}_i \notin \sigma} p_i=0\}.
\end{equation}
Here $\sigma \in \Delta$ means that
$\sigma$ is a cone in $\Delta$,
and $\vec{n}_i \in \sigma$ means that
$\vec{n}_i$ is a generator of some
one-dimensional cone in $\sigma$.

\noindent
3. $k$ vectors $\vec{n}_1,\vec{n}_2,\ldots,\vec{n}_k$
in ${\bf R}^d$ satisfy $(k-d)$ relations
\begin{equation}
\sum_{i=1}^k Q_i^{(a)} \vec{n}_i=0
\label{eq:Q-n}
\end{equation}
with $a=1,2,\ldots,k-d$.
Then the action of $({\bf C}^*)^{k-d}$
on $p_i$ is defined as
\begin{equation}
p_i \rightarrow \lambda_1^{Q_i^{(1)}}
\lambda_2^{Q_i^{(2)}} \cdots
\lambda_{k-d}^{Q_i^{(k-d)}} p_i, \quad
\lambda_a \in {\bf C}^*.
\end{equation}

An important point is that a set of vectors
$\{\vec{n}_1,\ldots ,\vec{n}_k\}$
determines the action of $({\bf C}^*)^{k-d}$
on ${\bf C}^k$,
but does not determine $F_\Delta$.
To determine $F_\Delta$
we must specify which vectors generate
each cone $\sigma$ in $\Delta$.
The specification is called a triangulation
of $\Delta$.
In general there are various triangulations
for the same set of vectors
$\{\vec{n}_1,\ldots ,\vec{n}_k\}$.
Different triangulations correspond to
different $F_\Delta$
and hence different toric varieties $V_\Delta$.

Various properties of a toric variety $V_\Delta$
can be expressed in terms of its fan $\Delta$.
For example, $V_\Delta$ is compact if and only if
$\Delta$ spans the whole of ${\bf R}^d$.

What we want to investigate is geometry in the
neighborhood of singularities in Calabi-Yau manifolds,
so we consider cases where $V_\Delta$ has
zero Chern class.
For these cases
a fan $\Delta$ must have the following property,
\begin{equation}
\vec{\mu} \cdot \vec{n}_i=1 \quad
\rm{for} \; \forall i
\label{eq:c_1}
\end{equation}
for a certain vector $\vec{\mu} \in {\bf Z}^d$.
This condition leads to relations among $({\bf C}^*)^{k-d}$
charges,
\begin{equation}
\sum_i Q^{(a)}_i=0
\end{equation}
due to (\ref{eq:Q-n}).
(\ref{eq:c_1}) also implies that
all the points specified by vectors
${\vec{n}_1,\ldots ,\vec{n}_k}$ lie in a
certain hyperplane $\Gamma$.
Thus we can represent a fan $\Delta$
in terms of the intersection of $\Delta$ with the
hyperplane $\Gamma$.
We call it as a toric diagram.

In the following we consider cases that the fan $\Delta$
is simplicial, for which $V_\Delta$ has at most quotient
singularities.
Simplicial fan is such that each cone $\sigma$
can be written in the form
\begin{equation}
\sigma={\bf R}_{\geq 0} \, \vec{n}_1 + \cdots +
{\bf R}_{\geq 0} \, \vec{n}_r
\end{equation}
for some linearly independent vectors
${\vec{n}_1,\ldots ,\vec{n}_r \in {\bf Z}^d}$.
When $r=d$, we define a volume for simplicial
cones to be $d!$ times the volume of the
polyhedron with vertices
$O,\vec{n}_1,\ldots ,\vec{n}_d$.
Here we take $\vec{n}_i$ to be the first
nonzero lattice point on the ray
${\bf R}_{\geq 0} \, \vec{n}_i$
\footnote{For fans satisfying (\ref{eq:c_1}),
the volume of a cone can be represented by
"area" of intersection of the cone with the
hyperplane $\Gamma$.
Unit of the area is defined as the area
of the intersection between $\Gamma$ and
a cone with unit volume.}.
$V_\Delta$ is smooth if every $d$-dimensional
cone in the fan has volume one.
If some cones have volume greater than one,
$V_\Delta$ has quotient singularities.
However we can obtain a smooth toric variety by
subdividing the cone until each cone has
volume one.
This procedure corresponds to a blow up of
the singularity.
An important point is that the resolution
of the singularity is not necessarily unique.
There are often numerous ways of subdividing
the cones in $\Delta$.
Thus there are numerous smooth varieties
that can arise from different ways of resolving
the singularity.
In general these varieties are topologically
distinct.
For Calabi-Yau threefold such topologically
distinct varieties can be connected by flops.

\begin{figure}[hbt]
\begin{center}
\leavevmode
\epsfbox{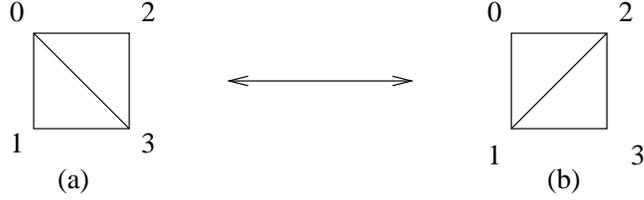}
\caption{The toric diagrams which represent a flop.
\label{fig:toricflop}}
\end{center}
\end{figure}

A flop is a transformation of a manifold into
a topologically different manifold which replaces
some homologically nontrivial cycle
${\bf CP}^1$ with another ${\bf CP}^1$.
In toric language, a flop is described by
toric diagrams which include figure \ref{fig:toricflop}.
We denote toric varieties corresponding
figure \ref{fig:toricflop}(a)
and \ref{fig:toricflop}(b)
as $V_{\Delta_a}$ and $V_{\Delta_b}$ respectively.
The toric variety $V_{\Delta_a}$ is written as
\begin{equation}
V_{\Delta_a}=({\bf C}^4-F_{\Delta_a})/{\bf C}^*
\end{equation}
where $F_{\Delta_a}=\{p_1=p_2=0\}$ and ${\bf C}^*$
charge of $p_i$ are $Q=(1,-1,-1,1)$ due to
the relation
$\vec{n}_0-\vec{n}_1-\vec{n}_2+\vec{n}_3=0$.
The toric variety $V_{\Delta_b}$ is
\begin{equation}
V_{\Delta_b}=({\bf C}^4-F_{\Delta_b})/{\bf C}^*
\end{equation}
where $F_{\Delta_b}=\{p_0=p_3=0\}$ and ${\bf C}^*$
charge is the same as that for $\Delta_a$.

To see the difference between the
two manifolds, we define subsets of ${\bf C}^4$ as
$U_a=\{p_0=p_3=0,(p_1,p_2) \neq (0,0)\}$,
$U_b=\{p_1=p_2=0,(p_0,p_3) \neq (0,0)\}$ and
$\tilde U={\bf C}^4-U_a \cup U_b \cup O$.
Then the toric varieties are written as
$V_{\Delta_a}=(\tilde U \cup U_a)/{\bf C}^*,
V_{\Delta_b}=(\tilde U \cup U_b)/{\bf C}^*$.
Here $\tilde U/{\bf C}^*$ is common to both manifolds.
Different contributions come from $U_a/{\bf C}^*$ and
$U_a/{\bf C}^*$.
$U_a/{\bf C}^*=\{(p_1,p_2) \in {\bf C}^2;
(p_1,p_2) \neq (0,0)\}/{\bf C}^*$
is a ${\bf CP}^1$, which we call ${\bf CP}^1_a$,
and $U_b/{\bf C}^*=\{(p_0,p_3) \in {\bf C}^2;
(p_0,p_3) \neq (0,0)\}/{\bf C}^*$
is another  ${\bf CP}^1$, which we call ${\bf CP}^1_b$.
Topology change occurs by
the replacement of ${\bf CP}^1_a$ by ${\bf CP}^1_b$.

The quotient construction of a toric variety
can be physically realized as a classical
vacuum of $N=2$ supersymmetric two dimensional
gauged linear sigma model\cite{Witten}.
In this approach, $p_i$ is the lowest component of
a chiral superfield with $U(1)$ charge $Q_i$.
The quotient by ${\bf C}^* \simeq {\bf R} \times S^1$
may be viewed to having taken place in two stages.
First we fix the ${\bf R}$ degree of freedom by imposing
D-flatness condition and then divide by
$U(1)$ gauge symmetry.
The most important point is that inclusion of
Fayet-Iliopoulos D-term with coefficient $\xi$
provides the degrees of freedom which are the
counterparts to the possible triangulations of
$\Delta$.

In the gauged linear sigma model description,
a flop is described by a model with the following
D-flatness condition
\begin{equation}
|p_0|^2-|p_1|^2-|p_2|^2+|p_3|^2=\xi.
\end{equation}
The $U(1)$ action on four chiral superfields is
$p_0 \rightarrow \lambda p_0$,
$p_1 \rightarrow \lambda^{-1} p_1$,
$p_2 \rightarrow \lambda^{-1} p_2$,
$p_3 \rightarrow \lambda p_3$.
For $\xi<0$, the vacuum moduli space specified by
the D-flatness condition do not contain $\{p_1=p_2=0\}$.
This condition is equivalent to removing the 
point set $F_{\Delta_a}$ in toric description.
For $\xi>0$, the vacuum moduli space do not
contain $\{p_0=p_3=0\}$.
This condition is equivalent to removing the 
point set $F_{\Delta_b}$.
For both cases $|\xi|$ controls the size of
${\bf CP}^1$ associated with the blow up.
The flop occurs in passing from $\xi<0$ to $\xi>0$
and vice versa.
Thus the sign of the Fayet-Iliopoulos parameter
specifies the triangulation as noted above.
Classically there is a singularity at $\xi=0$,
but as long as theta angle is generic,
smooth topology change can occur
\cite{Witten}.

\section{D-branes on ${\bf C}^3/{\bf Z}_n$}
\reseteqnum

We consider a typeII string theory on an
orbifold ${\bf C}^3/{\bf Z}_n$,
with ${\bf Z}_n$ a subgroup of $SU(3)$.
We begin with several remarks on orbifolds itself.
We take complex coordinates for ${\bf C}^3$
as $X^\mu$ $(\mu=1,2,3)$,
and an action of a generator $g$ of ${\bf Z}_n$
on $X^\mu$ as
\begin{equation}
g:X^\mu \rightarrow \omega^{a_\mu} X^\mu,
\quad \omega = {\rm exp}(2\pi i/n)
\label{eq:orbifold}
\end{equation}
where $(a_1,a_2,a_3)$ are integers which
satisfy $a_1+a_2+a_3 \simeq 0$, mod $n$.
Here we consider cases $a_\mu \neq 0$
since an orbifold with $a_\mu=0$ for some $\mu$
is a direct product of
${\bf C}^2/{\bf Z}_n$ and ${\bf C}$.
Thus orbifolds are labeled by $n$ and
$\vec{a}=(a_1,a_2,a_3)$,
but different $\vec{a}$'s often give the same
orbifold.
That is, if there is some integer $k$ which
has no common factor with $n$ except one
and satisfies the equation
\begin{equation}
(\omega^{a_1},\omega^{a_2},\omega^{a_3})
=((\omega^k)^{a'_1},(\omega^k)^{a'_2},(\omega^k)^{a'_3}),
\label{eq:equivalence}
\end{equation}
the orbifold with
$\vec{a}=(a_1,a_2,a_3)$ is the same as
the orbifold with $\vec{a}=(a'_1,a'_2,a'_3)$.
This is because the two cases give the same
action of ${\bf Z}_n$ on $X^\mu$ if we take
all elements $g^l$, ($l=0,1,\ldots ,n-1$)
into account.
It includes a trivial case
$a_\mu \rightarrow a_\mu+n$ $(k=1)$.
Apart from this,
an exchange between $a_\mu$ and $a_\nu$ does not
change the geometry since it is merely
an exchange between $X^\mu$ and $X^\nu$.
We present explicit examples of the
equivalence in the following sections.

To describe D-branes on the orbifold
${\bf C}^3/{\bf Z}_n$,
we first consider $n$ D-branes on ${\bf C}^3$.
It gives four dimensional $N=4$ supersymmetric
$U(n)$ gauge theory.
Bosonic field contents of the theory are
a gauge field $A_\alpha$, ($\alpha=0,1,2,3$),
and three complex scalar fields $X^\mu$,
($\mu=1,2,3$) where
$(X^\mu)^\dagger = X^{\bar{\mu}}$.
We then impose invariance under ${\bf Z}_n$.
It acts on the space-time indices of ${\bf C}^3$
as (\ref{eq:orbifold})
\footnote{Note that the resulting theory
has $N=1$ supersymmetry due to the relation
$a_1+a_2+a_3 \simeq 0$, mod $n$.},
and on Chan-Paton indices
(which label the D-branes)
in the regular representation
$\gamma(g)_{ij}=\delta_{ij} \omega^i$.
Thus the generator $g$ of ${\bf Z}_n$
acts on the gauge field as
\begin{equation}
g:A_\alpha \rightarrow
\gamma (g) A_\alpha \gamma (g)^{-1}.
\end{equation}
Invariance under ${\bf Z}_n$ gives
$A_{\alpha \, ij}=A_{\alpha \, i} \delta_{ij}$,
where $i,j=0,1,\ldots,n-1$,
therefore the gauge group $U(n)$ is reduced
to $U(1)^n$.

For $X^\mu$, ${\bf Z}_n$ acts as
\begin{equation}
g:X^\mu \rightarrow
\omega^{a_\mu} \gamma (g) X^\mu \gamma (g)^{-1}.
\end{equation}
Invariance under ${\bf Z}_n$ gives
\begin{equation}
X^\mu_{ij}=X^\mu_{ij} \delta_{j,i+a_\mu},
\end{equation}
which leave $3n$ fields.
We denote these fields as
\begin{equation}
x_i=X^1_{i,i+a_1}, \quad
y_i=X^2_{i,i+a_2}, \quad
z_i=X^3_{i,i+a_3}.
\end{equation}
Since all fields have charge zero for diagonal
$U(1)$, nontrivial gauge symmetry is $U(1)^{n-1}$.

The field contents can be described
by quiver diagrams\cite{DM,JM}.
It consists of $n$ vertices associated with
gauge fields $A_{\alpha \, i}$
and $3n$ oriented links associated with
$X^\mu_{ij}$.
The links represent the information on
$U(1)^n$ charges of $X^\mu_{ij}$.
The equivalence (\ref{eq:equivalence}) 
except $k=1$ corresponds to a certain
exchange among vertices.

To obtain vacuum moduli space of the theory,
we first impose F-flatness conditions.
F-flatness conditions give $(2n-2)$ equations
\begin{equation}
x_i z_{i+a_1}=z_i x_{i+a_3}, \quad
y_i z_{i+a_2}=z_i y_{i+a_3},
\end{equation}
so the resulting space is (n+2)-dimensional,
which we denote as ${\cal N}$.
Then we impose D-flatness conditions and divide
by the gauge symmetry.
D-flatness conditions give (real) $(n-1)$ equations
\begin{equation}
|x_{i-a_1}|^2+|y_{i-a_2}|^2+|z_{i-a_3}|^2
-|x_i|^2-|y_i|^2-|z_i|^2=\zeta_i,
\end{equation}
where $i=1,2,\ldots,n-1$ and $\zeta_i$ is a
coefficient of Fayet-Iliopoulos D-term,
which comes from the twisted sector of closed strings.
Together with a gauge fixing of $U(1)^{n-1}$
symmetry, we have three dimensional space
as expected.
This is the vacuum moduli space of the
D-brane world-volume gauge theory.
We denote it by ${\cal M}$.

Note that although F-flatness condition leads to
$[X^\mu, X^\nu]=0$, $[X^\mu, X^{\bar{\mu}}]$
does not vanish if $\zeta_i \neq 0$.
Thus the vacuum moduli space is embedded nontrivially
into a larger non-commuting configuration space.

We express the geometry of the moduli space
by using toric method as follows.
First, we express $(n+2)$-dimensional space
${\cal N}$ by a toric variety
$V_{\Delta_{\cal N}}=
({\bf C}^{n+2+k}-F_{\Delta_{\cal N}})/({\bf C}^*)^k$.
To make the translation into a toric description,
we use another definition\cite{AGM} of toric varieties
from that described in section 2.
First, we introduce $3n$ variables $u_i$ and $(n+2)$
variables $v_i$ as $(u_1,\ldots ,u_{3n})
=(x_1,\ldots ,x_{n-1},y_1,\ldots ,y_{n-1},
x_0,y_0,z_1,\ldots ,z_{n-1},z_0)$,
$(v_1,\ldots ,v_{n+2})
=(x_0,y_0,z_1,\ldots ,z_{n-1},z_0)$.
Then the solution of the F-flatness conditions
is written in the form
\begin{equation}
u_i=\prod_{j=1}^{n+2} v_j^{m_{ij}}
\end{equation}
where $m_{ij} \in {\bf Z}$.
We can regard $m_{ij}$ as the $j$-th component of
the $i$-th $(n+2)$-dimensional vector $\vec{m}_i$.
These $3n$ vectors define a cone $\hat \sigma$
in ${\bf R}^{n+2}$ as
\begin{equation}
\hat \sigma=\{\vec{m} \in {\bf R}^{n+2};
        \vec{m}=\sum_{i=1}^{3n}
        a_i \vec{m}_i, \quad a_i>0\}.
\end{equation}
We now define dual cone $\sigma$ of
$\hat \sigma$ by
\begin{equation}
\sigma=\{\vec{n} \in {\bf R}^{n+2};
             \vec{m} \cdot \vec{n} \geq 0,
             \quad \forall \vec{m} \in \hat \sigma\}.
\end{equation}
A collection of cones, $\sigma$ and its faces,
defines the fan $\Delta_{\cal N}$.

Then we express the three dimensional vacuum moduli
space ${\cal M}$ by a toric variety.
It is obtained from ${\cal N}$ by imposing $(n-1)$
D-flatness conditions and $U(1)^{n-1}$ gauge fixing.
It is equivalent to the holomorphic quotient
by $({\bf C}^*)^{n-1}$ after removing an appropriate
point set.
Together with the holomorphic quotient by
$({\bf C}^*)^k$ in ${\cal N}$, ${\cal M}$ is expressed as
${\cal M}=({\bf C}^{n+2+k}
-F_{\Delta_{\cal M}})/({\bf C}^*)^{k+n-1}$.
Note that $({\bf C}^*)^{n-1}$ charges of
$p_0, p_1, \ldots, p_{n+1+k}$
are determined from $U(1)^{n-1}$ charges of
$(x_i,y_i,z_i)$ through the following relations
\begin{equation}
u_i=\prod p_j^{{\vec{n}_j} \cdot {\vec{m}_i}}.
\end{equation}

In fact there are redundancy in the
homogeneous coordinates,
so we can eliminate such coordinates by using
appropriate $({\bf C}^*)$ quotients.

\section{Orbifold resolution and topology change}
\reseteqnum

In this section, we study the vacuum moduli
space of D-brane world-volume gauge theory
on ${\bf C}^3/{\bf Z}_n$ for
$n=3,5,7,9,11$.
Although $n=3,5$ cases were already
investigated in \cite{DGM},
we describe the results for these cases
rather elaborately
to explain the method of analysis.

\subsection{D-branes on ${\bf C}^3/{\bf Z}_3$}

For ${\bf C}^3/{\bf Z}_3$, any model is equivalent to
the model with $\vec{a}=(2,2,-1)$
\footnote{In the following, we take $a_3=-1$
for convenience.
This choice is always allowed
due to the equivalence under (\ref{eq:equivalence})
and $a_{\mu} \leftrightarrow a_{\nu}$.}.
For example, a model with $\vec{a}=(1,1,1)$
is equivalent to the model with
$\vec{a}=(2,2,-1)$ as follows,
\begin{equation}
(\omega^1,\omega^1,\omega^1)
=(\omega^{1+3},\omega^{1+3},\omega^{1-3})
=((\omega^2)^2,(\omega^2)^2,(\omega^2)^{-1}).
\end{equation}

In this case, the cone $\hat \sigma$
is generated by nine vectors,
\begin{eqnarray}
&&\vec{m}_1 = (1,0,1,0,-1),\nonumber\\
&&\vec{m}_2 = (1,0,0,1,-1),\nonumber\\
&&\vec{m}_3 = (0,1,1,0,-1),\nonumber\\
&&\vec{m}_4 = (0,1,0,1,-1),\nonumber\\
&&\vec{m}_5 = (1,0,0,0, 0),\\
&&\vec{m}_6 = (0,1,0,0, 0),\nonumber\\
&&\vec{m}_7 = (0,0,1,0, 0),\nonumber\\
&&\vec{m}_8 = (0,0,0,1, 0),\nonumber\\
&&\vec{m}_9 = (0,0,0,0, 1).\nonumber
\end{eqnarray}

The fan $\Delta_{\cal N}$ is determined by
the condition $\vec{m}_i \cdot \vec{n} \geq 0$
$(i=1,2,\ldots,9)$,
and its one dimensional cones are generated by
\begin{eqnarray}
&&\vec{n}_0=(1,0,0,0,0),\nonumber\\
&&\vec{n}_1=(0,1,0,0,0),\nonumber\\
&&\vec{n}_2=(0,0,1,0,0),\\
&&\vec{n}_3=(0,0,0,1,0),\nonumber\\
&&\vec{n}_4=(1,1,0,0,1),\nonumber\\
&&\vec{n}_5=(0,0,1,1,1).\nonumber
\end{eqnarray}
They satisfy the relation
$\vec{n}_0+\vec{n}_1-\vec{n}_2
-\vec{n}_3-\vec{n}_4+\vec{n}_5=0$,
so the ${\bf C}^*$ charges of $p_i$'s are given by
\begin{equation}
Q_{\cal N}=\left(
\begin{array}{cccccc}
 1& 1&-1&-1&-1& 1\\
\end{array}
\right).
\end{equation}
Combining $({\bf C}^*)^2$ charges
which originate from $U(1)^2$ gauge group
of the original D-brane world-volume theory,
we have $({\bf C}^*)^3$ charges of $p_i$'s as
\begin{equation}
Q_{\rm tot}=\left(
\begin{array}{cccccc}
 1& 1&-1&-1&-1& 1\\
 0& 0&-1& 1& 0& 0\\
 0& 0& 0&-1& 1& 0
\end{array}
\right).
\label{eq:Qtot}
\end{equation}
The resulting toric variety takes the form
$({\bf C}^6-F_{\Delta})/({\bf C}^*)^3$.
The charge matrix (\ref{eq:Qtot})
determines six vectors which generate
one dimensional cones in $\Delta$.
\begin{eqnarray}
&&\vec{n}_0=(-1,-1,1),\nonumber\\
&&\vec{n}_1=(0,1,1),\nonumber\\
&&\vec{n}_2=(0,0,1),\\
&&\vec{n}_3=(0,0,1),\nonumber\\
&&\vec{n}_4=(0,0,1),\nonumber\\
&&\vec{n}_5=(1,0,1).\nonumber
\end{eqnarray}
Here $\vec{n}_2=\vec{n}_3=\vec{n}_4$,
so two of the three homogeneous coordinates
$p_2,p_3,p_4$ are redundant.
It implies that the toric variety is written
in the form
\begin{equation}
({\bf C}^6-F_{\Delta})/({\bf C}^*)^3
=(({\bf C}^4-F_{\Delta_{\cal M}}) \times ({\bf C}^*)^2)
/({\bf C}^*)^3
=({\bf C}^4-F_{\Delta_{\cal M}})/{\bf C}^*.
\end{equation}
This is the moduli space ${\cal M}$
of the D-brane world-volume gauge theory.

\begin{figure}[hbt]
\begin{center}
\leavevmode
\epsfxsize=3cm
\epsfbox{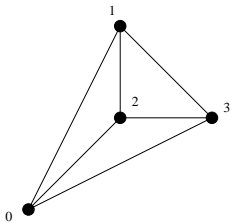}
\caption{The toric diagram for $n=3$, $\vec{a}=(2,2,-1)$.
\label{fig:z3}}
\end{center}
\end{figure}

To determine the triangulation of $\Delta_{\cal M}$,
we go to the gauged linear sigma model description.
It is a $U(1)^3$ gauge theory with six chiral
superfields.
The D-flatness conditions are
\begin{equation}
|p_0|^2+|p_1|^2-|p_2|^2
-|p_3|^2-|p_4|^2+|p_5|^2=0,
\end{equation}
\begin{equation}
-|p_2|^2+|p_3|^2=\zeta_1,
\end{equation}
\begin{equation}
-|p_3|^2+|p_4|^2=\zeta_2.
\end{equation}
Here $\zeta_1$ and $\zeta_2$ are coefficients
of Fayet-Iliopoulos D-term associated to the
$U(1)^2$ gauge group of the original D-brane theory.
Now we eliminate two of $p_2$, $p_3$ and $p_4$.
When $\zeta_1<0$ and $\zeta_2>0$,
$p_2$ and $p_4$ can not vanish,
so they take value in $({\bf C}^*)^2$.
Then we can eliminate these fields
by using D-flatness conditions and
$U(1)^2$ gauge symmetry.
Thus we have
\begin{equation}
|p_0|^2+|p_1|^2-3|p_2|^2+|p_5|^2
=-\zeta_1+\zeta_2 \equiv \xi.
\end{equation}
It is the D-flatness condition of a $U(1)$
gauged linear sigma model with four
chiral superfields with charges
\begin{equation}
Q_{\cal M}=\left(
\begin{array}{ccccc}
 1& 1&-3& 1& -\zeta_1+\zeta_2.
\end{array}
\right)
\end{equation}
Here we include the information on the
Fayet-Iliopoulos D-term parameter.
As we are considering the region
$\zeta_1<0$ and $\zeta_2>0$,
the Fayet-Iliopoulos parameter of the
resulting $U(1)$ gauge theory
$\xi=-\zeta_1+\zeta_2$ is positive.
Then $(p_0,p_1,p_5) \neq (0,0,0)$,
which implies $F_\Delta = \{p_0=p_1=p_5=0\}$
in the toric description.
The toric diagram which gives this $F_\Delta$
is figure \ref{fig:z3}.
The diagram consists of three triangles
with area one, so the corresponding toric
variety is a smooth manifold obtained by
blowing up the orbifold singularity of
${\bf C}^3/{\bf Z}_3$.

In \cite{DGM}, it was also shown that
the result does not depend on
which set of fields are eliminated
out of $p_2$, $p_3$ and $p_4$.
Therefore the orbifold phase is projected out.

\subsection{D-branes on ${\bf C}^3/{\bf Z}_5$}

\begin{figure}[hbt]
\begin{center}
\leavevmode
\epsfxsize=11cm
\epsfbox{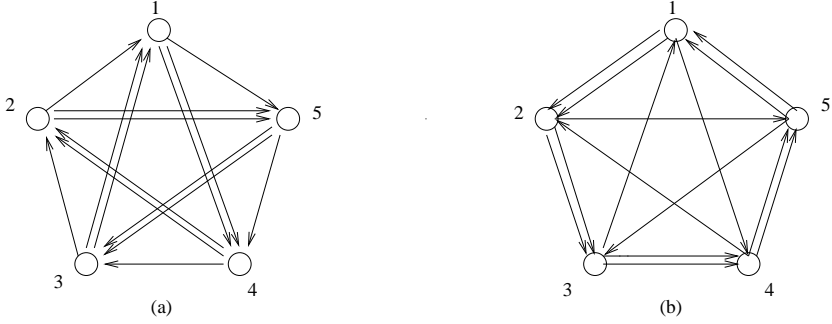}
\caption{(a)The quiver diagram for $n=5$,
$\vec{a}=(3,3,-1)$,
(b) the quiver diagram for $n=5$, $\vec{a}=(1,1,3)$.
The two diagrams are related by a permutation
of vertices.\label{fig:q5}}
\end{center}
\end{figure}

For ${\bf C}^3/{\bf Z}_5$, any model is equivalent to
the model with $\vec{a}=(3,3,-1)$.
For example, a model with $\vec{a}=(1,1,3)$
is equivalent to the model with
$\vec{a}=(3,3,-1)$ as follows,
\begin{equation}
(\omega^1,\omega^1,\omega^3)
=(\omega^{1+5},\omega^{1+5},\omega^{3-5})
=((\omega^2)^3,(\omega^2)^3,(\omega^2)^{-1})
\end{equation}
The quiver diagram for $\vec{a}=(1,1,3)$
and that for $\vec{a}=(3,3,-1)$
are related by the following permutation
$\tau$ of vertices
(see figure \ref{fig:q5}),
\begin{equation}
\tau=
\left(
\begin{array}{ccccc}
1&2&3&4&5\\
1&3&5&2&4
\end{array}
\right).
\end{equation}

\begin{figure}[hbt]
\begin{center}
\leavevmode
\epsfxsize=4.5cm
\epsfbox{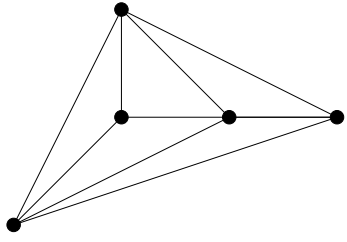}
\caption{The toric diagram for $n=5$, $\vec{a}=(3,3,-1)$.
\label{fig:z5}}
\end{center}
\end{figure}

The moduli space takes the form
$({\bf C}^{13}-F_{\Delta})/({\bf C}^*)^{10}
=({\bf C}^{5}-F_{\Delta_{\cal M}})/({\bf C}^*)^{2}$,
and the charge matrix is given by
\begin{equation}
Q_{\cal M}=\left(
\begin{array}{ccccc}
 1& 1&-3& 1& 0 \\
 0& 0& 1&-2& 1
\end{array}
\right).
\end{equation}
Figure \ref{fig:z5} represents the corresponding
toric diagram.

\subsection{D-branes on ${\bf C}^3/{\bf Z}_7$}

\begin{figure}[hbt]
\begin{center}
\leavevmode
\epsfxsize=11cm
\epsfbox{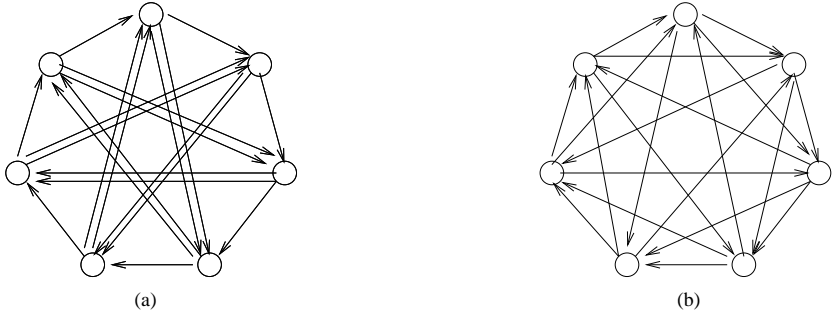}
\caption{(a) The quiver diagram for $n=7$,
$\vec{a}=(4,4,-1)$,
(b) the quiver diagram for $n=7$, $\vec{a}=(3,5,-1)$.
\label{fig:q7}}
\end{center}
\end{figure}

There are two types of models for ${\bf C}^3/{\bf Z}_7$.
One has $\vec{a}=(4,4,-1)$ and the other has
$\vec{a}=(3,5,-1)$.
The quiver diagrams are depicted in figure
\ref{fig:q7}.

\begin{figure}[hbt]
\begin{center}
\leavevmode
\epsfxsize=6cm
\epsfbox{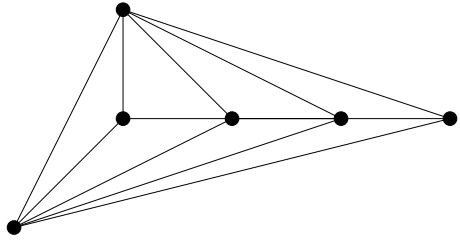}
\caption{The toric diagram for $n=7$, 
$\vec{a}=(4,4,-1)$.
\label{fig:z7a}}
\end{center}
\end{figure}

For the case of $\vec{a}=(4,4,-1)$,
one can see that $\Delta_{\cal N}$ is generated by
31 vectors in ${\bf R}^9$, and hence ${\cal N}$
takes the form $({\bf C}^{31}-F_{\Delta_{\cal N}})/
({\bf C}^*)^{22}$.
Combining $({\bf C}^*)^6$ quotient coming from
$U(1)^6$ gauge symmetry of the D-brane theory,
${\cal M}$ takes the form
$({\bf C}^{31}-F_{\Delta})/({\bf C}^*)^{28}$.
However one can see that 25 vectors out of 31 are redundant.
Hence we eliminate 25 coordinates by $({\bf C}^*)^{25}$
action and finally the moduli space takes the form
$({\bf C}^{6}-F_{\Delta_{\cal M}})/({\bf C}^*)^{3}$.
For one choice of redundant variables,
the charge matrix is given by
\begin{equation}
Q_{\cal M}=\left(
\begin{array}{ccccccc}
 1& 1&-3& 1& 0& 0& 2\zeta_1+\zeta_2+\zeta_3+\zeta_4 \\
 0& 0& 1&-2& 1& 0& \zeta_2+\zeta_5 \\
 0& 0& 0& 1&-2& 1& \zeta_3+\zeta_6
\end{array}
\right).
\end{equation}
For this choice of redundant variables,
Fayet-Iliopoulos parameters must satisfy
\begin{eqnarray}
&&\zeta_1>0, \quad \zeta_2>0, \quad \zeta_3>0, \quad
  \zeta_2+\zeta_5>0, \quad \zeta_3+\zeta_6>0, \quad
  \zeta_1+\zeta_2+\zeta_3+\zeta_4>0, \nonumber\\
&&\zeta_1+\zeta_2+\zeta_3+\zeta_4+\zeta_5>0, \quad
  \zeta_1+\zeta_2+\zeta_3+\zeta_4+\zeta_5+\zeta_6>0.
\end{eqnarray}
Under these conditions,
Fayet-Iliopoulos parameters of the resulting
$U(1)^3$ gauged linear sigma model satisfy
inequalities
\begin{eqnarray}
&&\xi_1=2\zeta_1+\zeta_2+\zeta_3+\zeta_4>0, \nonumber\\
&&\xi_2=\zeta_2+\zeta_5>0, \\
&&\xi_3=\zeta_3+\zeta_6>0. \nonumber
\end{eqnarray}
These inequalities determine the phase uniquely.
The corresponding toric diagram is figure
\ref{fig:z7a}.

\begin{figure}[hbt]
\begin{center}
\leavevmode
\epsfxsize=4.5cm
\epsfbox{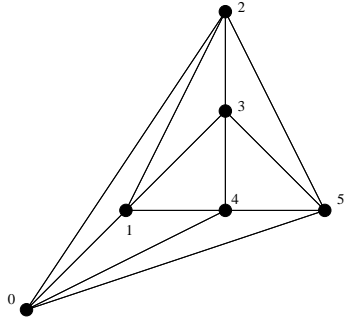}
\caption{The toric diagram for $n=7$, 
$\vec{a}=(3,5,-1)$.\label{fig:z7b}}
\end{center}
\end{figure}

For the case of $\vec{a}=(3,5,-1)$,
one can see that $\Delta_{\cal N}$ is generated by
24 vectors in ${\bf R}^9$, and hence ${\cal N}$
takes the form $({\bf C}^{24}-F_{\Delta_{\cal N}})/
({\bf C}^*)^{15}$.
Combining the $({\bf C}^*)^6$ coming from
$U(1)^6$ gauge symmetry of the D-brane theory,
${\cal M}$ takes the form
$({\bf C}^{24}-F_{\Delta})/({\bf C}^*)^{21}$.
However one can see that 18 vectors out of 24 are redundant.
Hence the moduli space takes the form
$({\bf C}^{6}-F_{\Delta_{\cal M}})/({\bf C}^*)^{3}$.
For one choice of redundant variables,
the charge matrix is given by
\begin{equation}
Q_{\cal M}=\left(
\begin{array}{ccccccc}
 1& 0&-2& 1& 0& 0& \zeta_1+\zeta_2+\zeta_3 \\
 0& 1& 0&-2& 1& 0& \zeta_1+\zeta_4+\zeta_5 \\
 0& 0& 1& 0&-2& 1& \zeta_2+\zeta_4+\zeta_6
\end{array}
\right).
\end{equation}
It determines eight vectors which generate
one dimensional cones for $\Delta_{\cal M}$,
\begin{eqnarray}
&&\vec{n}_0=(1,2,1),\nonumber\\
&&\vec{n}_1=(2,0,1),\nonumber\\
&&\vec{n}_2=(1,1,1),\\
&&\vec{n}_3=(1,0,1),\nonumber\\
&&\vec{n}_4=(0,0,1),\nonumber\\
&&\vec{n}_5=(-1,-1,1).\nonumber
\end{eqnarray}
For the choice of redundant variables
Fayet-Iliopoulos parameters must satisfy
\begin{eqnarray}
&&\zeta_1>0, \quad \zeta_2>0, \quad \zeta_4>0, \quad
  \zeta_1+\zeta_2+\zeta_3>0, \quad
  \zeta_1+\zeta_4+\zeta_5>0,\nonumber\\
&&\zeta_2+\zeta_4+\zeta_6>0,\quad
  \zeta_1+\zeta_2+\zeta_3+\zeta_4+\zeta_5>0, \quad
  \zeta_1+\zeta_2+\zeta_3+\zeta_4+\zeta_6>0,\\
&&\zeta_1+\zeta_2+\zeta_4+\zeta_5+\zeta_6>0, \quad
  \zeta_1+\zeta_2+\zeta_3+\zeta_4+\zeta_5+\zeta_6>0.
  \nonumber
\end{eqnarray}
Under these conditions,
Fayet-Iliopoulos parameters of the resulting
$U(1)^3$ gauged linear sigma model satisfy
inequalities
\begin{eqnarray}
&&\xi_1=\zeta_1+\zeta_2+\zeta_3>0, \nonumber\\
&&\xi_2=\zeta_1+\zeta_4+\zeta_5>0, \\
&&\xi_3=\zeta_2+\zeta_4+\zeta_6>0. \nonumber
\end{eqnarray}
The triangulation of the toric diagram
is uniquely determined as in
figure \ref{fig:z7b}.
Thus only the completely resolved phase is realized.

Here we comment on the correspondence between the classification
of models by $\vec{a}$ and the classification of toric diagrams.
In the $n=7$ case, if we allow $a_\mu=0$ for some $\mu$,
there are three types of models, 
$\vec{a}=(4,4,-1)$, $\vec{a}=(3,5,-1)$ and $\vec{a}=(1,6,0)$.
On the other hand, triangles with area seven
whose corners lie on the lattice ${\bf Z}^2$
are also classified into three types;
the first is the triangle corresponding to figure \ref{fig:z7a},
the second is that corresponding to figure \ref{fig:z7b}
and the third is that with corners $(0,0,1)$, $(0,1,1)$
and $(7,0,1)$.
Here the classification of triangles is defined
by the following equivalence relation:
two triangles are equivalent if they are related
by $GL(2,{\bf Z})$ transformation with determinant
$\pm 1$.
In fact, there is a relation between $\vec{a}$ and
the triangle which appears in the corresponding toric
diagram\cite{Aspinwall}.
That is, if we write three corners of the triangle
as $\vec{l}_\mu=(l_{\mu 1},l_{\mu 2},l_{\mu 3})$,
$(\mu=1,2,3)$,
the following relations
\begin{equation}
\frac{1}{n} \sum_{\mu =1}^3 a_{\mu} l_{\mu i}
\in {\bf Z}
\end{equation}
hold for an appropriate order of $(a_1,a_2,a_3)$.
Note that the triangle corresponding to the
$\vec{a}=(1,6,0)$ model have lattice points
on its codimension one boundary.
This is a characteristic feature to the cases
whose singularity is non-isolated (see section 5).

{}From the examples treated in this paper,
it seems that the classification of models on
${\bf C}^3/{\bf Z}_n$ corresponds to the
classification of triangles with area $n$
(see section 6).

\subsection{D-branes on ${\bf C}^3/{\bf Z}_9$}

For ${\bf C}^3/{\bf Z}_9$,
there are two types of models.
One has $\vec{a}=(5,5,-1)$ and the other has
$\vec{a}=(4,6,-1)$
\footnote{$\vec{a}=(3,3,3)$ also satisfies
$a_1+a_2+a_3 \simeq 0$ modulo 9,
but it is nothing but the model on ${\bf C}^3/{\bf Z}_3$
with $\vec{a}=(1,1,1)$ .}.

For the $\vec{a}=(4,6,-1)$ case,
the action of ${\bf Z}_9$ on $X^\mu$ is
\begin{equation}
(X^1,X^2,X^3) \rightarrow
(\omega^{4k} X^1,\omega^{6k} X^2,\omega^{-k} X^3),
\quad \omega^9=1
\end{equation}
where $k=1,\ldots ,9$.
For $k=3$,
$(X^1,X^2,X^3) \rightarrow
(\omega^3 X^1,X^2,\omega^{-3} X^3)$,
which means that there is a non-trivial
fixed point $(0,X^2,0)$ with $X^2 \neq 0$.
Therefore the corresponding space has
a non-isolated singularity.
We discuss such cases in section
\ref{sec:nonisolated}.

For the case of $\vec{a}=(5,5,-1)$,
one can see that $\Delta_{\cal N}$ is generated by
78 vectors in ${\bf R}^{11}$, and hence ${\cal N}$
takes the form $({\bf C}^{78}-F_{\Delta_{\cal N}})/
({\bf C}^*)^{67}$.
Combining the $({\bf C}^*)^8$ quotient coming from
$U(1)^8$ gauge symmetry of the D-brane theory,
${\cal M}$ takes the form
$({\bf C}^{78}-F_{\Delta})/({\bf C}^*)^{75}$.
However one can see that 71 vectors out of 78 are redundant.
Hence the moduli space takes the form
$({\bf C}^{7}-F_{\Delta_{\cal M}})/({\bf C}^*)^{4}$.
For one choice of redundant variables,
the charge matrix is given by
\begin{equation}
Q_{\cal M}=\left(
\begin{array}{cccccccc}
1& 1&-3& 1& 0& 0& 0 \\
0& 0& 1&-2& 1& 0& 0 \\
0& 0& 0& 1&-2& 1& 0 \\
0& 0& 0& 0& 1&-2& 1
\end{array}
\right)
\end{equation}

\begin{figure}[hbt]
\begin{center}
\leavevmode
\epsfxsize=7.5cm
\epsfbox{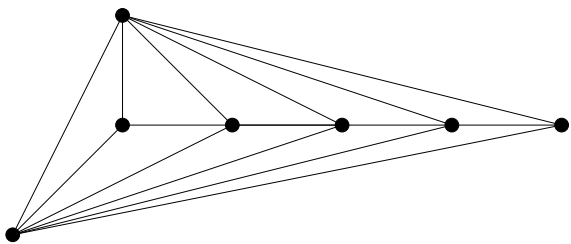}
\caption{The toric diagram for $n=9$, 
$\vec{a}=(5,5,-1)$.\label{fig:z9a}}
\end{center}
\end{figure}

By a similar analysis to the previous sections
we can see that the toric diagram is
given by figure \ref{fig:z9a}.

\subsection{D-branes on ${\bf C}^3/{\bf Z}_{11}$
and topology change}

For ${\bf C}^3/{\bf Z}_{11}$, there are two types of models.
One has $\vec{a}=(6,6,-1)$ and the other has
$\vec{a}=(5,7,-1)$.

\begin{figure}[hbt]
\begin{center}
\leavevmode
\epsfxsize=9cm
\epsfbox{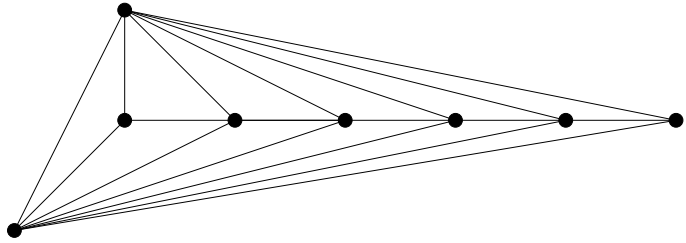}
\caption{The toric diagram for $n=11$, 
$\vec{a}=(6,6,-1)$.\label{fig:z11a}}
\end{center}
\end{figure}

For the case of $\vec{a}=(6,6,-1)$,
one can see that $\Delta_{\cal N}$ is generated by
201 vectors in ${\bf R}^{13}$, and hence ${\cal N}$
takes the form $({\bf C}^{201}-F_{\Delta_{\cal N}})/
({\bf C}^*)^{188}$.
Combining the $({\bf C}^*)^{10}$ quotient coming from
$U(1)^{10}$ gauge symmetry of the D-brane theory,
${\cal M}$ takes the form
$({\bf C}^{201}-F_{\Delta})/({\bf C}^*)^{198}$.
However one can see that 193 vectors out of 201 are redundant.
Hence the moduli space takes the form
$({\bf C}^{8}-F_{\Delta_{\cal M}})/({\bf C}^*)^{5}$
with the charge matrix,
\begin{equation}
Q_{\cal M}=\left(
\begin{array}{ccccccccc}
 1& 1&-3& 1& 0& 0& 0& 0 \\
 0& 0& 1&-2& 1& 0& 0& 0 \\
 0& 0& 0& 1&-2& 1& 0& 0 \\
 0& 0& 0& 0& 1&-2& 1& 0 \\
 0& 0& 0& 0& 0& 1&-2& 1
\end{array}
\right).
\end{equation}
The corresponding toric diagram
is figure \ref{fig:z11a}.

\begin{figure}[hbt]
\begin{center}
\leavevmode
\epsfxsize=6cm
\epsfbox{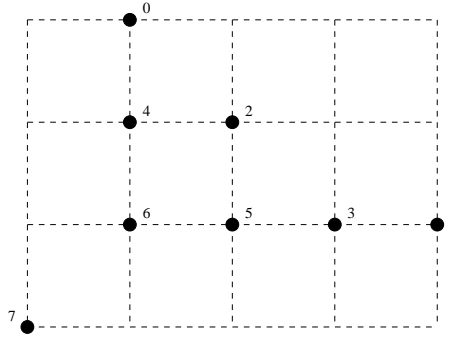}
\caption{Eight vectors generating one dimensional cones
in $\Delta_{\cal M}$
for the $n=11$, $\vec{a}=(5,7,-1)$ model.
\label{fig:z11b}}
\end{center}
\end{figure}

For the case of $\vec{a}=(5,7,-1)$,
one can see that $\Delta_{\cal N}$ is generated by
91 vectors in ${\bf R}^{13}$, and hence ${\cal N}$
takes the form $({\bf C}^{91}-F_{\Delta_{\cal N}})/
({\bf C}^*)^{78}$.
Combining the $({\bf C}^*)^{10}$ quotient coming from
$U(1)^{10}$ gauge symmetry of the D-brane theory,
${\cal M}$ takes the form
$({\bf C}^{91}-F_{\Delta})/({\bf C}^*)^{88}$.
However one can see that 83 vectors out of 91 are redundant.
Hence we can eliminate 83 coordinates by $({\bf C}^*)^{83}$
action and finally the moduli space takes the form
$({\bf C}^{8}-F_{\Delta_{\cal M}})/({\bf C}^*)^{5}$.
For one choice of redundant variables,
the charge matrix is given by
\begin{equation}
Q_{\cal M}=\left(
\begin{array}{ccccccccc}
 1& 0&-2& 1& 0& 0& 0& 0&
  \zeta_1+\zeta_2+\zeta_3+\zeta_4+\zeta_5 \\
 0& 1& 0&-2& 0& 1& 0& 0&
  \zeta_1+\zeta_6+\zeta_7 \\
 1& 0& 0& 0&-2& 0& 1& 0&
  \zeta_1+\zeta_4+\zeta_5+\zeta_8 \\
 0& 0& 0& 1& 0&-2& 1& 0&
  \zeta_2+\zeta_3+\zeta_8 \\
 0& 0& 1& 0& 0& 0&-2& 1&
  \zeta_2+\zeta_4+\zeta_6+\zeta_8+2\zeta_9+\zeta_{10}
\end{array}
\right).
\end{equation}
It determines eight vectors which generate
one dimensional cones in $\Delta_{\cal M}$,
\begin{eqnarray}
&&\vec{n}_0=(0,2,1),\nonumber\\
&&\vec{n}_1=(3,0,1),\nonumber\\
&&\vec{n}_2=(1,1,1),\nonumber\\
&&\vec{n}_3=(2,0,1),\\
&&\vec{n}_4=(0,1,1),\nonumber\\
&&\vec{n}_5=(1,0,1),\nonumber\\
&&\vec{n}_6=(0,0,1),\nonumber\\
&&\vec{n}_7=(-1,-1,1).\nonumber
\end{eqnarray}
Figure \ref{fig:z11b} represents these eight vectors.

For the choice of redundant variables
Fayet-Iliopoulos parameters must satisfy
\begin{eqnarray}
&&\zeta_1 > 0, \quad
  \zeta_2 > 0, \quad
  \zeta_4 > 0, \quad
  \zeta_6 > 0, \quad
  \zeta_9 > 0, \quad
  \zeta_2 +\zeta_8 > 0, \quad
  \zeta_4 +\zeta_8 > 0, \nonumber\\
&&\zeta_1 +\zeta_2 +\zeta_3 > 0, \quad
  \zeta_1 +\zeta_6 +\zeta_7 > 0, \quad
  \zeta_2 +\zeta_3 +\zeta_8 > 0, \quad
  \zeta_1 +\zeta_4 +\zeta_5 +\zeta_8 > 0,
  \nonumber\\
&&\zeta_1 +\zeta_2 +\zeta_3 +\zeta_4 +\zeta_5 > 0, \quad
  \zeta_2 +\zeta_4 +\zeta_6 +\zeta_9 +\zeta_{10} > 0,
  \nonumber\\
&&\zeta_2 +\zeta_4 +\zeta_6 +\zeta_8 +\zeta_9
  +\zeta_{10} > 0,\quad
  \zeta_1 +\zeta_2 +\zeta_3 +\zeta_4 +\zeta_5
  +\zeta_6 +\zeta_7 > 0, \nonumber\\
&&\zeta_2 +\zeta_3 +\zeta_4 +\zeta_6 +\zeta_8
  +\zeta_9 +\zeta_{10} > 0, \quad
  \zeta_1 +\zeta_2 +\zeta_3 +\zeta_4 +\zeta_5
  +\zeta_6 +\zeta_7 +\zeta_8 > 0, \nonumber\\
&&\zeta_1 +\zeta_2 +\zeta_4 +\zeta_5 +\zeta_6
  +\zeta_8 +\zeta_9 +\zeta_{10} > 0,\\
&&\zeta_1 +\zeta_2 +\zeta_3 +\zeta_4 +\zeta_5
  +\zeta_6 +\zeta_8 +\zeta_9 +\zeta_{10} > 0,
  \nonumber\\
&&\zeta_1 +\zeta_2 +\zeta_3 +\zeta_4 +\zeta_6
  +\zeta_7 +\zeta_8 +\zeta_9 +\zeta_{10} > 0,
  \nonumber\\
&&\zeta_1 +\zeta_2 +\zeta_3 +\zeta_4 +\zeta_5 +\zeta_6
  +\zeta_7 +\zeta_8 +\zeta_9 +\zeta_{10} > 0.
  \nonumber
\end{eqnarray}
Under these conditions,
Fayet-Iliopoulos parameters of the resulting
$U(1)^5$ gauged linear sigma model satisfy
the following inequalities
\begin{eqnarray}
&&\xi_1=\zeta_1+\zeta_2+\zeta_3+\zeta_4
       +\zeta_5>0, \nonumber \\
&&\xi_2=\zeta_1+\zeta_6+\zeta_7>0, \nonumber \\
&&\xi_3=\zeta_1+\zeta_4+\zeta_5+\zeta_8>0,
\label{eq:zeta11b} \\
&&\xi_4=\zeta_2+\zeta_3+\zeta_8>0, \nonumber \\
&&\xi_5=\zeta_2+\zeta_4+\zeta_6+\zeta_8
       +2\zeta_9+\zeta_{10}>0, \nonumber
\end{eqnarray}

\begin{figure}[hbt]
\begin{center}
\leavevmode
\epsfxsize=12cm
\epsfbox{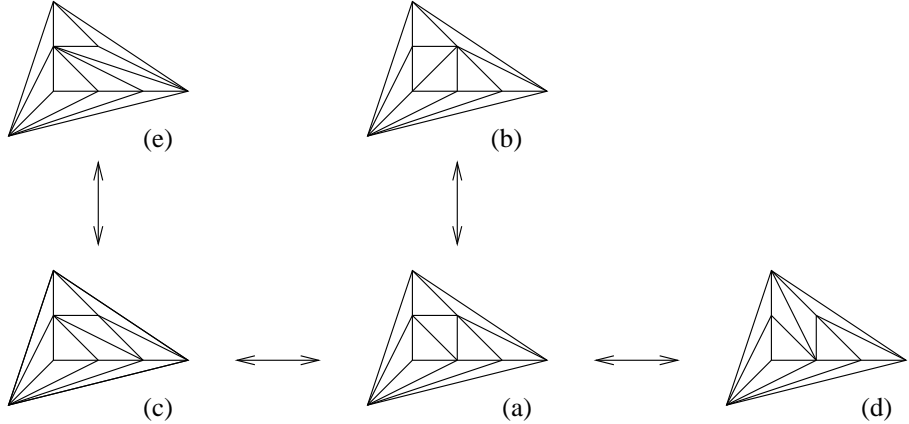}
\caption{The toric diagrams for the $n=11$,
$\vec{a}=(5,7,-1)$ model.
There are five phases connected by flops.
\label{fig:z11bflop}}
\end{center}
\end{figure}

However these conditions are not enough to determine the
phase. To see this we explicitly write down
the D-flatness conditions.
\begin{eqnarray}
|p_0|^2-2|p_2|^2+|p_3|^2=\xi_1, \nonumber \\
|p_1|^2-2|p_3|^2+|p_5|^2=\xi_2, \nonumber \\
|p_0|^2-2|p_4|^2+|p_6|^2=\xi_3, \\
|p_3|^2-2|p_5|^2+|p_6|^2=\xi_4, \nonumber \\
|p_2|^2-2|p_6|^2+|p_7|^2=\xi_5. \nonumber
\end{eqnarray}
Using these equations,
we have the following equations
\begin{eqnarray}
&&|p_2|^2-|p_4|^2-|p_5|^2+|p_6|^2
=(-\xi_1+\xi_3+\xi_4)/2 \equiv \eta_1, \nonumber \\
&&|p_3|^2-|p_2|^2-|p_5|^2+|p_4|^2
=(\xi_1-\xi_3+\xi_4)/2 \equiv \eta_2, \\
&&|p_0|^2-|p_2|^2-|p_4|^2+|p_5|^2
=(\xi_1+\xi_3-\xi_4)/2 \equiv \eta_3, \nonumber \\
&&|p_1|^2-|p_2|^2-|p_3|^2+|p_4|^2
=(\xi_1+2\xi_2-\xi_3+\xi_4)/2
\equiv \eta_4. \nonumber
\end{eqnarray}
If $\eta_i$ can have both positive and negative value,
topology change occurs as the sign of $\eta_i$
changes.
In fact, we can see that it occurs as follows.
Under the conditions (\ref{eq:zeta11b}),
$\eta_i$'s must satisfy
\begin{eqnarray}
&&\eta_1+\eta_2 = \xi_4 > 0, \nonumber \\
&&\eta_1+\eta_3 = \xi_3 > 0, \nonumber \\
&&\eta_2+\eta_3 = \xi_1 > 0, \\
&&\eta_3+\eta_4 = \xi_1+\xi_2 > 0, \nonumber \\
&&\eta_4-\eta_2 = \xi_2 > 0. \nonumber
\end{eqnarray}
These inequalities restrict possible sign of $\eta_i$
and the following five cases, which we call
(a), (b), (c), (d) and (e), are allowed.
\begin{equation}
(\eta_1,\eta_2,\eta_3,\eta_4) \sim
\left\{
\begin{array}{cc}
(+,+,+,+),& \qquad ({\rm a}) \\
(-,+,+,+),& \qquad ({\rm b}) \\
(+,-,+,+),& \qquad ({\rm c}) \\
(+,+,-,+),& \qquad ({\rm d}) \\
(+,-,+,-).& \qquad ({\rm e})
\end{array}
\right.
\end{equation}
We can see that the triangulation of the toric
diagram is uniquely determined for each case.
The corresponding toric diagrams are
depicted in figure \ref{fig:z11bflop}.
Each triangulation consists of eleven cones
with volume one, thus it represents a phase
of completely resolved manifolds
with definite topology.
The phases (a) and (b) are connected by
a flop with respect to the parallelogram
with vertices $\vec{n}_2$, $\vec{n}_4$, $\vec{n}_5$
and $\vec{n}_6$,
and the flop is accomplished by the change of the sign
of $\eta_1$.
Similarly the phases (a) and (c),
(a) and (d), (c) and (e) are connected by a flop
controlled by $\eta_2$,
$\eta_3$, $\eta_4$, respectively.

We can see that
there is no triangulation except these five cases
which completely resolves the orbifold singularity.

\section{D-branes on orbifolds with
non-isolated singularities
\label{sec:nonisolated}}
\reseteqnum
Orbifolds often have non-isolated singularities
as in the $n=9$, $\vec{a}=(4,6,-1)$ case.
Consider an orbifold ${\bf C}^3/{\bf Z}_n$ with
$\vec{a}=(a_1,a_2,a_3)$.
If $n$ and $a_\mu$ have a common factor except one,
the orbifold has a non-isolated fixed point of
${\bf Z}_n$ and hence
has a non-isolated singularity.
Thus there are models with non-isolated singularity
if $n$ is not a prime number.
Especially orbifold singularities are always
non-isolated if $n$ is even.
The analyses in the previous sections
can be almost straightforwardly applicable
to these cases.
We have explicitly calculated the moduli space of
D-branes on such orbifolds for $n=4,6$ cases
\footnote{In the $n=2$ case, $a_\mu$ must be
zero (mod 2) for some $\mu$,
so the orbifold is a direct product of
${\bf C}^2/{\bf Z}_2$ and ${\bf C}$.},
and $n=9$, $\vec{a}=(4,6,-1)$ case
and verified that only the geometric phases
are realized.

\begin{figure}[hbt]
\begin{center}
\leavevmode
\epsfxsize=12cm
\epsfbox{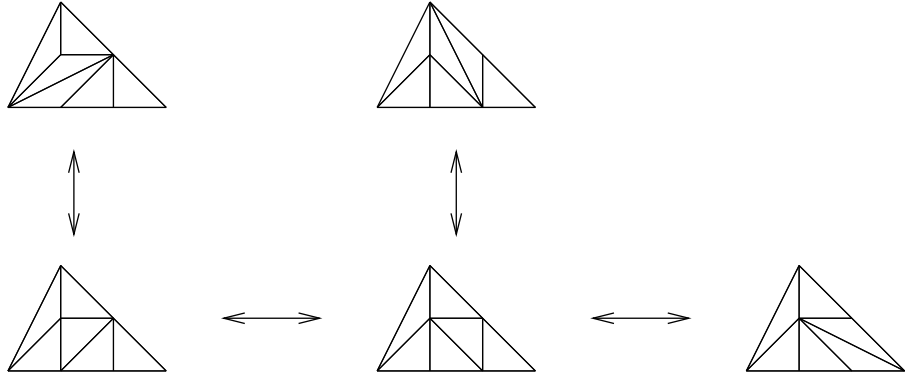}
\caption{The toric diagrams for the $n=6$,
$\vec{a}=(2,5,-1)$ model.
There are five phases connected by flops.
\label{fig:z6bflop}}
\end{center}
\end{figure}

Topologically distinct phases emerge
in the $n=6$, $\vec{a}=(2,5,-1)$ case,
in which there are five phases connected by flops
(figure \ref{fig:z6bflop}).
We have also found that the moduli space
consists of six blown-up phases connected by flops
in the $n=9$, $\vec{a}=(4,6,-1)$ case.
We can see that for both cases
there is no other triangulation
which completely resolves the orbifold singularity.

We also note that toric diagrams corresponding to
non-isolated cases always have vertices on its codimension
one boundary.

\section{Discussion}
\reseteqnum

From the analyses we have made in this paper,
it seems that there is a correspondence between
the classification of models on ${\bf C}^3/{\bf Z}_n$
by $\vec{a}$ and the classification of triangles
with area $n$ (as noted in subsection 4.3),
and what phases are allowed for each model is
determined according to the method
of subdivisions of the triangle with area $n$
into $n$ triangles with area one.
(Here vertices of each triangle lie on the
lattice ${\bf Z}^2$.)
It means that only completely blown-up phases are realized
as pointed out in \cite{DGM},
and topology change can occur
if the triangle with area $n$
includes parallelogram with respect to
the lattice ${\bf Z}^2$.

It will be interesting to prove
the above-mentioned rule on the allowed phases in general.

\vskip 1cm
\centerline{\large\bf Acknowledgements}

I would like to thank T.~Aoyama
for valuable discussions.

\newcommand{\J}[4]{{\sl #1} {\bf #2}(19#3) #4}
\newcommand{\NP}{Nucl.~Phys.}
\newcommand{\PL}{Phys.~Lett.}
\newcommand{\PR}{Phys.~Rev.}
\newcommand{\PRL}{Phys.~Rev.~Lett.}
\newcommand{\MPL}{Mod.~Phys.~Lett.}
\newcommand{\PTP}{Prog.~Theor.~Phys.}
\newcommand{\CMP}{Comm.~Math.~Phys.}
\newcommand{\PRep}{Phys.~Rep.}

\end{document}